\documentclass{optica-article}

\journal{opticajournal} 

\articletype{Research Article}

\usepackage{lineno}
\usepackage{multirow}
\usepackage{gensymb}

\begin{document}

\title{Brillouin nonlinearity characterizations of a high refractive index silicon oxynitride platform}

\author{Kaixuan Ye,\authormark{1} Akshay Keloth,\authormark{1} Yvan Klaver,\authormark{1} Alessio Baldazzi,\authormark{2} Gioele Piccoli,\authormark{3} Matteo Sanna,\authormark{2} Lorenzo Pavesi,\authormark{2} Mher Ghulinyan,\authormark{3} and David Marpaung\authormark{1,*}}

\address{\authormark{1}Nonlinear Nanophotonics Group, MESA+ Institute of Nanotechnology, University of Twente, Enschede, Netherlands\\
\authormark{2}Department of Physics, University of Trento, Trento, Italy\\
\authormark{3}Sensors and Devices, Fondazione Bruno Kessler, Trento, Italy}

\email{\authormark{*}david.marpaung@utwente.nl}

\begin{abstract} 
Silicon oxynitride (SiON) is a low-loss and versatile material for linear and nonlinear photonics applications. Controlling the oxygen-to-nitrogen (O/N) ratio in SiON provides an effective way to engineer its optical and mechanical properties, making it a great platform for the investigation of on-chip optomechanical interactions, especially the stimulated Brillouin scattering (SBS). Here we report the Brillouin nonlinearity characterization of a SiON platform with a specific O/N ratio (characterized by a refractive index of $n=1.65$). First, we introduce this particular SiON platform with fabrication details. Subsequently, we discuss various techniques for the on-chip Brillouin nonlinearity characterizations. In particular, we focus on the intensity-modulated pump-probe lock-in amplifier technique, which enables ultra-sensitive characterization. Finally, we analyze the Brillouin nonlinearities of this SiON platform and compare them with other SiON platforms. This work underscores the potential of SiON for on-chip Brillouin-based applications. Moreover, it paves the way for Brillouin nonlinearity characterization across various material platforms.
\end{abstract}

\section{Introduction}
Silicon oxynitride (SiON), an amorphous material with a refractive index between silicon oxide ($n=1.45$) and silicon nitride ($n=2.0$), emerges as a promising integrated platform for both linear and nonlinear photonic applications \cite{Piccoli2022}. It has a broad transparency window spanning from visible spectrum \cite{Jung2022} to near-infrared range \cite{Bernard2021}. Additionally, it features optical waveguides with low propagation losses \cite{Chen2017}, negligible multi-photon absorption, and low tensile stress in films up to a few micrometers thick \cite{Moss2013}.   

By manipulating the oxygen to nitrogen (O/N) stoichiometric ratio in SiON, its optical properties, including both the refractive index $n$ and the Kerr nonlinear index $n_2$, can be continuously tuned. In the low refractive index regime (silica-like), these SiON waveguides exhibit reduced propagation losses, making them advantageous for applications in biosensing \cite{Samusenko2016} and passive photonic circuits, such as ultra-high-Q cavities \cite{chen2017chip}, taper designs \cite{Jia2023}, filters and interferometers \cite{melloni2007progress}. Conversely, SiON waveguides with high refractive index (nitride-like) exhibit a more pronounced Kerr effect, making them highly useful in optical frequency comb-based applications \cite{Xu2021, Reimer2016} and in on-chip entangled photon-pair generation schemes \cite{Piccoli2022}.

Adjusting the O/N ratio in SiON not only modifies its optical properties but also affects its mechanical properties. The tunability in both optical and mechanical properties makes SiON an interesting material for the investigation of optomechanical interactions, especially the stimulated Brillouin scattering (SBS). SiON waveguides with different O/N ratios exhibit distinct Brillouin nonlinearities. In \cite{Ye2023}, SiON waveguides with a relatively low refractive index ($n=1.51$), show a Brillouin gain coefficient of 0.32 m$^{-1}$W$^{-1}$ at 14.48~GHz. Very recently, SiON waveguides with high refractive index ($n=1.70$) have also been explored \cite{Zerbib2023}, showing a Brillouin gain coefficient of 0.09 m$^{-1}$W$^{-1}$ at 16~GHz. 

In this study, we report the Brillouin nonlinearity characterization of SiON waveguides with a different material composition, i.e., with a refractive index $n=1.65$. This characterization is enabled by the ultra-sensitive intensity-modulated pump-probe lock-in amplifier setup. We analyze the influence of various experimental settings, including modulation depth, modulation frequency, DC bias points, and detection frequency on the measurement results. With the optimized configurations, we successfully determined the Brillouin nonlinearity of the SiON waveguides under test, showing a Brillouin gain coefficient of 0.75 m$^{-1}$W$^{-1}$ with a Brillouin frequency shift of 17.15~GHz. Our results complement previous characterizations of SiON waveguides with different refractive indexes. Moreover, the analysis of the intensity-modulated pump-probe lock-in amplifier setup extends beyond SiON, facilitating the Brillouin nonlinearity characterization across diverse material platforms.

\section{SiON platform}
\begin{figure*}[b!]
\centering
\includegraphics[width=0.9\linewidth]{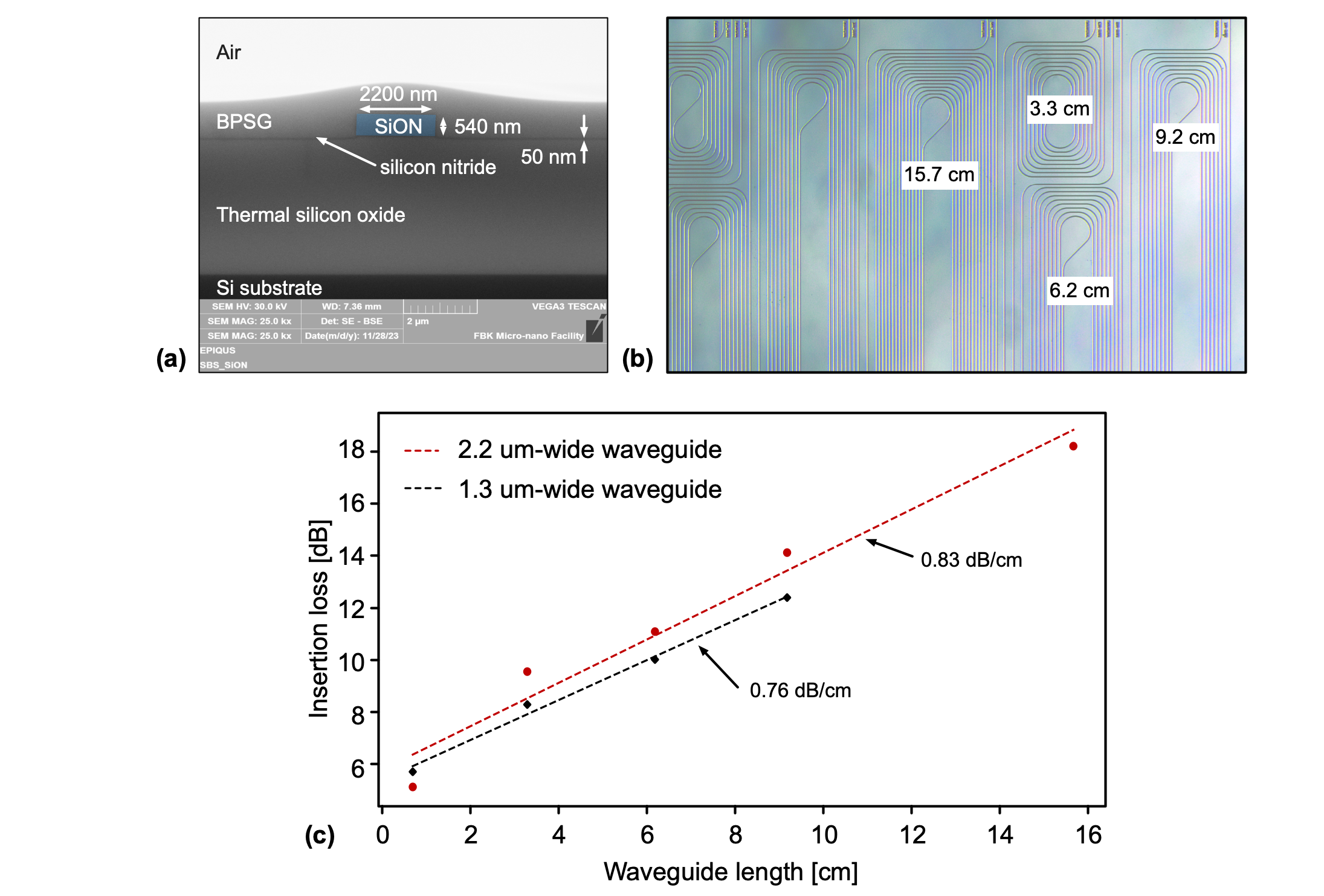}
\caption{(a) SEM image (with false color overlaid on the waveguide area) showing the cross section of the sample; (b) Microscope image of the top view of the sample; (c) Measured insertion loss of SiON waveguides in different lengths.}
\label{fig:chip}
\end{figure*}

The fabrication procedure of the SiON platform investigated in this work is described as follows. We start the fabrication process by growing a 3~$\mu$m silicon oxide layer with wet thermal oxidation on a silicon wafer. To complete the bottom cladding, an additional 1~$\mu$m Tetraethyl Orthosilicate (TEOS)  oxide was deposited using low-pressure chemical vapor deposition (LPCVD). The whole stack was annealed at 1050$\degree$C for a couple of hours to improve the cladding quality. The core SiON layer was deposited with a low-temperature plasma-enhanced chemical vapor deposition (PECVD) machine using silane, ammonia, and nitrous oxide precursor gases. An i-line stepper lithography was used to pattern the waveguides and to transfer them to the SiON layer using reactive ion etching (RIE). The so-made waveguides were treated at a temperature of 1050$\degree$C for 90 minutes in N$_2$ atmosphere to release the H$_2$ bonds to improve the optical properties. The waveguides were capped by a 50 nm thick LPCVD Si$_3$N$_4$ film to reduce the sidewall roughness thus improving the propagation loss. Finally, the waveguides were covered with 1150 nm thick borophosphosilicate glass (BPSG) via LPCVD and re-flowed at 925$\degree$C for partial planarization. Images of the fabricated device cross-section and the chip are shown in Fig. \ref{fig:chip} (a) and (b), respectively. 

We used a cut-back method to measure the coupling and propagation loss of the waveguides. Lensed fiber with 2~$\mu$m spot size is used to couple light into waveguide spirals of different lengths. The coupling loss is 2.7 and 2.9~dB/facet for 1.3~$\mu$m and 2.2~$\mu$m wide waveguides respectively. Additionally, the measured propagation losses are 0.76~dB/cm and 0.83~dB/cm respectively, as shown in Fig. \ref{fig:chip} (c).

\section{Brillouin nonlinearity characterization}
Accurate Brillouin nonlinearity characterization of SiON waveguides requires an ultra-sensitive setup, given the relatively small SBS gain. For justification, consider 1~W pump power launched into a 10~cm SiON waveguide. Taking into account a propagation loss of $\alpha=0.8$~dB/cm and a coupling loss of 3~dB/facet, the estimated SBS gain from the waveguide is in the range of 0.1~dB, a magnitude easily susceptible to noise fluctuations.

\begin{figure*}[t!]
\centering
\includegraphics[width=\linewidth]{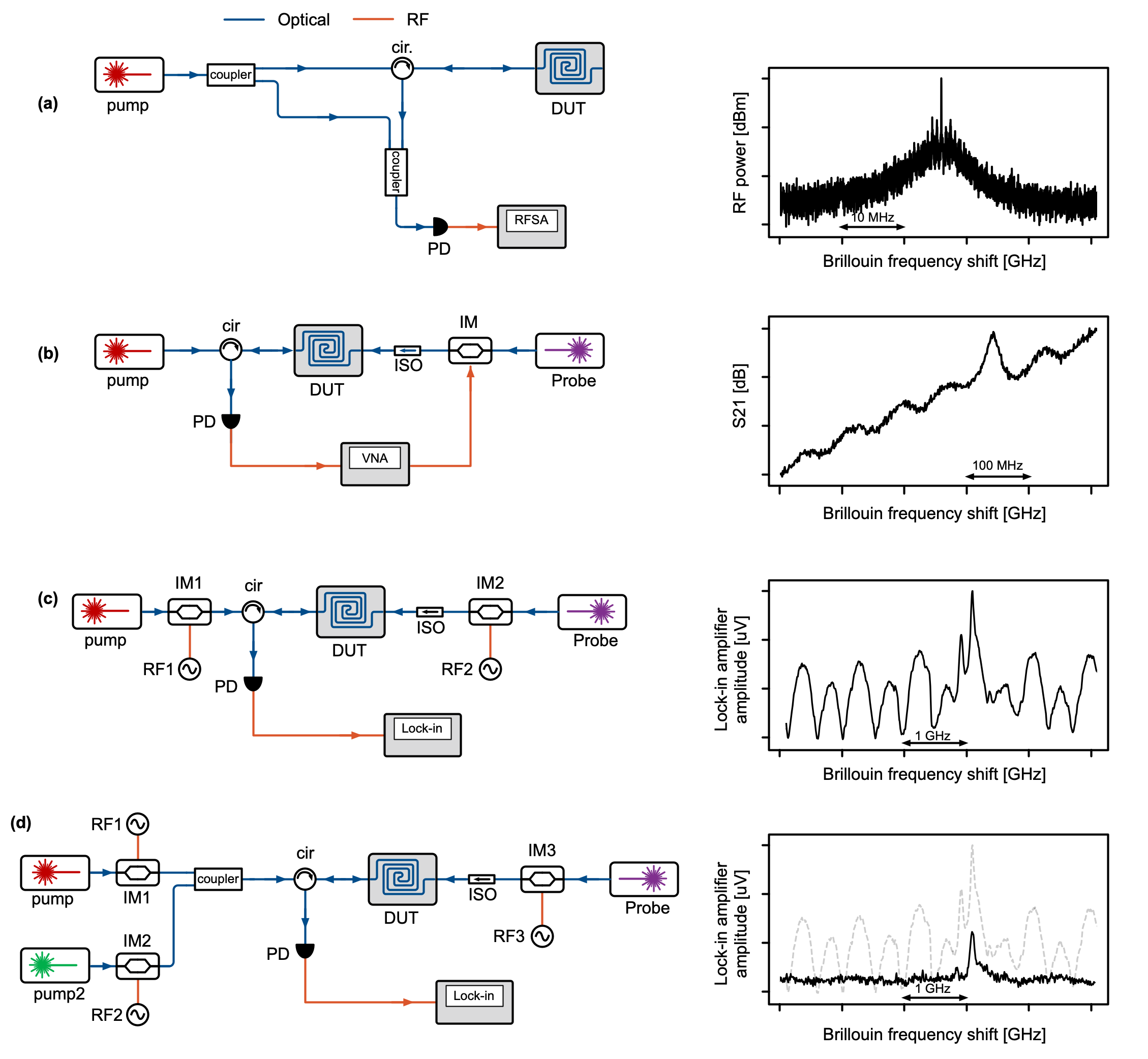}
\caption{Simplified on-chip Brillouin nonlinearity characterization setups and typical responses. (a) Single-pump RFSA-based heterodyne measurement; (b) Pump-probe sideband-sweeping VNA measurement; (c) Double-intensity-modulated pump-probe lock-in amplifier measurement; (d) Triple-intensity-modulated pump-probe lock-in amplifier measurement. Typical responses in the right panel only serve for illustration purposes and do not represent the real measurement results. RFSA: radio-frequency spectrum analyzer; VNA: vector-network analyzer.}
\label{fig:char}
\end{figure*}

Thus far, various techniques have been developed to measure on-chip Brillouin gain. In \cite{Kittlaus2016, Zerbib2023, Zhang2022}, a single-pump setup is applied, as shown in Fig.~\ref{fig:char}~(a). The thermally-generated Brillouin scattering signal is mixed with a local oscillator (derived from the same laser as the pump) for heterodyne detection with a radio-frequency spectrum analyzer (RFSA). Although straightforward, this configuration often demands high pump power to initiate the SBS process. 

To alleviate the requirement on the pump power, pump-probe techniques are more often applied for the on-chip Brillouin nonlinearity characterization \cite{Kittlaus2017,Gyger2020,Gundavarapu2019,Botter2022,Ye2023,Botter2023,Raphael2015-njp,Raphael2015}, as shown in Fig.~\ref{fig:char}~(b)-(d). The pump is usually put at a fixed frequency. Meanwhile, the probe frequency is swept across the Brillouin frequency shift region, either by electro-optical modulation or by controlling the current or temperature of the laser. Fig.~\ref{fig:char}~(b) shows the simplified schematic of the pump-probe setup based on the vector network analyzer (VNA). By sweeping the RF frequency applied to the modulator, the sideband of the probe will scan across the Brillouin frequency shift region of the waveguide. The sideband and carrier of the probe will beat at the photodiode (PD), and the Brillouin gain profile can be obtained \cite{neijts2023onchip, Pant2011}. This method enables accurate Brillouin frequency shift measurement. Nevertheless, fluctuations in the VNA would deteriorate the signal-to-noise (SNR) ratio, making it difficult for accurate readout when the SBS gain is small. 

To enhance the SNR of the detected signal, pump-probe setups based on the lock-in amplifier have also been developed \cite{Botter2022,Botter2023,Ye2023,Gyger2020}. The lock-in amplifier selectively detects signals pertaining to the SBS process, as a result, the sensitivity of this technique would be significantly higher than the above-mentioned techniques. Fig.~\ref{fig:char}~(c) shows the simplified schematic of the double-intensity-modulated pump-probe lock-in amplifier setup for the Brillouin nonlinearity characterization (the operating principles and the detailed setup will be included in later subsections). While it is effective in most cases \cite{Botter2022,Botter2023,Ye2023}, for samples with high Kerr nonlinearity or strong reflection at the facets, the oscillation in the noise floor would still obscure the signal of interest. To suppress the oscillation, a triple-intensity-modulated pump-probe lock-in amplifier setup, as shown in Fig.~\ref{fig:char}~(d), was also developed.

In this work, we characterized the Brillouin nonlinearity of the SiON waveguides with the intensity-modulated pump-probe lock-in amplifier technique. In the following subsections, we start by describing the operating principles of the double-intensity modulation setup. Subsequently, influences of various experimental parameters are investigated in detail. We provide guidelines for the optimized experimental configurations and justify the conditions when the triple-intensity modulation setup is necessary. In the end, we introduce the triple-intensity modulation setup applied in this work.

\subsection{Intensity-modulated pump-probe lock-in amplifier technique}
\begin{figure}[t!]
\centering
\includegraphics[width=\linewidth]{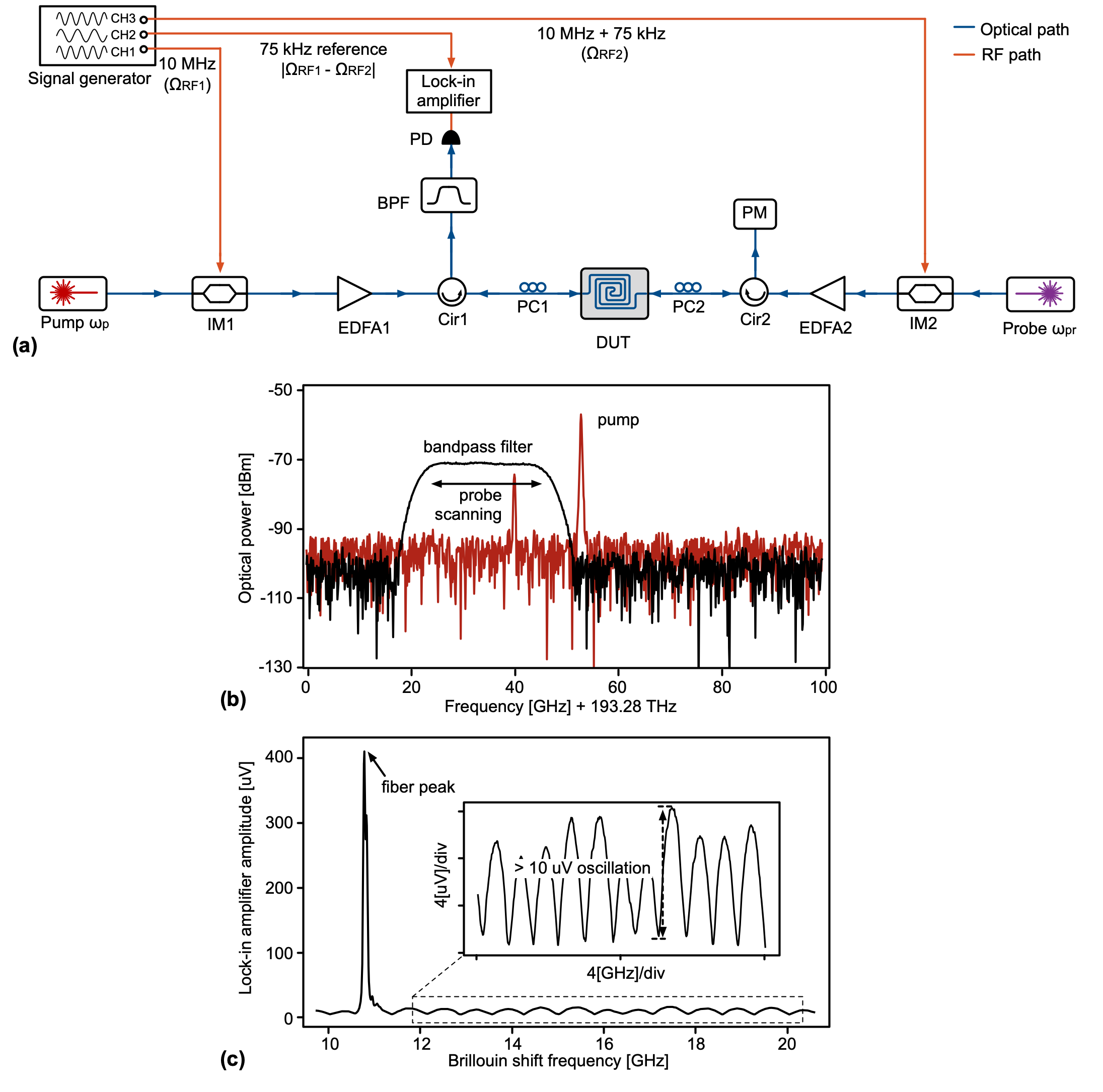}
\caption{(a) Schematic of the double-intensity-modulated lock-in amplifier setup; (b) Configurations of the pump, probe, and the optical bandpass filter; (c) The measured SBS signal from the lock-in amplifier. The fiber peak signal is clearly visible at the Brillouin frequency shift of 10.8~GHz, while the SBS signal from the waveguide is obscured within the fluctuating noise floor.}
\label{fig:double}
\end{figure}

Fig.~\ref{fig:double} (a) shows the schematic and the optical spectrum of the double-intensity-modulated pump-probe lock-in amplifier setup. The pump (1560.67~nm) and probe (frequency downshifted from the pump by tens of GHz) are coupled into the waveguide from opposite directions. The pump frequency is fixed while the probe frequency is swept by changing the current of the laser. Both the pump and probe are intensity-modulated with RF signals at slightly different frequencies, $\Omega_{RF1}$ (10~GHz) and $\Omega_{RF2}$ (10~GHz + 75~kHz). To achieve sufficient SBS gain, the pump is amplified with a high-power erbium-doped fiber amplifier (EDFA1). In the meantime, the probe is also amplified by EDFA2 to compensate for insertion losses and ensure adequate optical power at the photodiode. Two circulators (Cir1 and Cir2) are placed after the EDFAs to separate the transmitted probe from the pump and safeguard the probe laser against residual pump light. Additionally, polarization controllers (PC1 and PC2) are added to align the polarization of the pump and the probe. The transmitted probe, together with the reflected pump, comes out from Port 3 of Cir1. A bandpass filter is deployed to filter out the reflected pump and solely transmit the probe signal to the photodiode. Subsequently, the output of the photodiode is connected to a lock-in amplifier that detects signals at frequency $|\Omega_{RF1} - \Omega_{RF2}|$ (75~kHz). 

The optical spectrum configurations of the pump, probe, and bandpass filter are presented in Fig.~\ref{fig:double} (b). The pump frequency is fixed at the stopband of the bandpass filter, while the probe frequency is scanned within the passband of the filter. At each probe frequency, the lock-in amplifier collects the beating signal, which pertains to the SBS process. Having the probe frequency scanned across the Brillouin frequency shift region of both the fiber and the waveguide, the Brillouin gain profiles of both the fiber and the waveguide under test can be obtained.

In the experiment, the sweep speed of the probe must be slower than the time constant of the lowpass filter inside the lock-in amplifier for accurate data acquisition. Additionally, the step frequency of the sweeping needs to be sufficiently small to resolve the linewidth of the SBS responses of both the fiber and the waveguide.

In an ideal scenario, the lock-in amplifier measurement would display distinct SBS responses from both the fiber and the waveguide. Leveraging the well-studied properties of the single-mode fiber, the fiber peak in the experiment can serve as a reference for both locating the Brillouin frequency shift and calculating the Brillouin gain coefficient of the waveguide. Given the fiber peak and the waveguide peak reading from the lock-in amplifier is $V_{\rm fiber}$ and $V_{\rm wg}$, respectively, the Brillouin gain coefficient of the waveguide can be calculated as below \cite{Botter2022}:

\begin{equation}
g_{\rm{B}, \rm{wg}} = \frac{V_{\rm{wg}}}{V_{\rm{fiber}}}\frac{g_{\rm{B}, \rm{fiber}}L_{\rm{eff, fiber}}P_{\rm{pump,fiber}}}{L_{\rm{eff, wg}}P_{\rm{pump,wg}}}
\label{eq:comp}
\end{equation}

While Eq.~\eqref{eq:comp} provides a direct approach for estimating the Brillouin gain coefficient of the waveguide, it lacks consideration for the influence of experimental settings on measurement outcomes. Illustrated in Fig.~\ref{fig:double}~(c), sub-optimal experimental configurations would lead to oscillations in the noise floor, obscuring the SBS response of the waveguide even when the SBS response of the fiber is evident. This complication hinders the accurate estimation of the Brillouin gain coefficient of the waveguide using Eq.~\eqref{eq:comp}.

\subsection{SBS process in the intensity-modulated pump-probe lock-in amplifier setup}
To investigate the influences of various experimental settings, it is imperative to derive the SBS process with relevant parameters included. Assuming the pump is coupled from $z=0$ and probe is coupled from $z=L$, the SBS process in the waveguide is governed by \cite{Boyd2008}:

\begin{equation}
\begin{aligned}
-\frac{d P_{pr}}{d z} &= g_B P_{p} P_{pr} - \alpha P_{pr}\\
\frac{d P_{p}}{d z} &= -g_B P_{p} P_{pr} - \alpha P_{p},
\end{aligned}
\label{eq:SBSdiff}
\end{equation}
where $P_{pr}$ and $P_P$ is the intensity-modulated pump and probe signal, $g_B$ is the Brillouin gain coefficient, and $\alpha$ denotes the propagation loss.

The intensity-modulated pump and probe are influenced by the DC bias and the modulation depth, which can be expressed as:
\begin{subequations}
\begin{equation}
\begin{aligned}
&P_{p} \approx \frac{P_{p0}}{2} \left[1 - {\rm cos}\phi_{B1}J_0(m_1)+ 2{\rm sin}\phi_{B1}J_1(m_1){\rm cos}(\Omega_{RF1}t)\right]\\
\end{aligned}
\label{eq:power-pp}
\end{equation}
\begin{equation}
\begin{aligned}
&P_{pr} \approx \frac{P_{pr0}}{2} \left[1 - {\rm cos}\phi_{B2}J_0(m_2) + 2{\rm sin}\phi_{B2}J_1(m_2){\rm cos}(\Omega_{RF2}t)\right]
\end{aligned}
\label{eq:power-pr}
\end{equation}
\label{eq:power}
\end{subequations}
where $P_{p0}$ and $P_{pr0}$ denote the average power of the pump and probe. The term $\phi_B=\frac{\pi V_{bias}}{V_{\pi, DC}}$ signifies the phase shift induced by the DC bias, while $m=\frac{\pi V_{RF}}{V_{\pi,RF}}$ represents the modulation depth of the intensity modulator. The symbols $J_0$ and $J_1$ correspond to the zero- and first-order Bessel function applied to the modulation depth, i.e., $J = J(m)$.

The amplified probe can be calculated by applying expressions in \eqref{eq:power} into \eqref{eq:SBSdiff}. While the differential equations in \eqref{eq:SBSdiff} cannot be solved analytically, assuming the SBS gain is small and dropping the DC terms in \eqref{eq:power}, the transmitted probe power can be approximated as:

\begin{equation}
\begin{aligned}
    P_{pr(z=0)} 
    & \approx P_{pr0}{\rm sin}\phi_{B2}J_1(m_2){\rm cos}(\Omega_{RF2}t)e^{-\alpha L} \ \times \\
    &\left[1 + g_B P_{p0} {\rm sin}\phi_{B1}J_1(m_1)\int^L_0 {\rm cos}\left(\frac{2\pi}{\lambda_{\rm RF1}}z - \Omega_{RF1}t\right) e^{-\alpha z} dz \right]
\end{aligned}
\label{eq:trans-probe}
\end{equation}
where $\lambda_{\rm RF1} = \frac{2\pi c}{n_{g}\Omega_{RF1}}$ is the wavelength of the envelop of the intensity-modulated pump.

The second term in \eqref{eq:trans-probe} describes the SBS signal while considering the influence of various factors, including propagation loss, DC bias points, modulation depths, and modulation frequencies. This expression converges to a similar but simplified form as presented in \cite{Gyger2020} assuming that both intensity modulators are biased at the quadrature point, the modulation frequencies are low, and the propagation loss is negligible. However, it is crucial to note that these assumptions are generally not applicable in real scenarios, and all these parameters would influence the detected signal in the experiments.

\subsection{Influences of DC bias voltages and modulation depths}
\begin{figure}[t!]
\centering
\includegraphics[width=\linewidth]{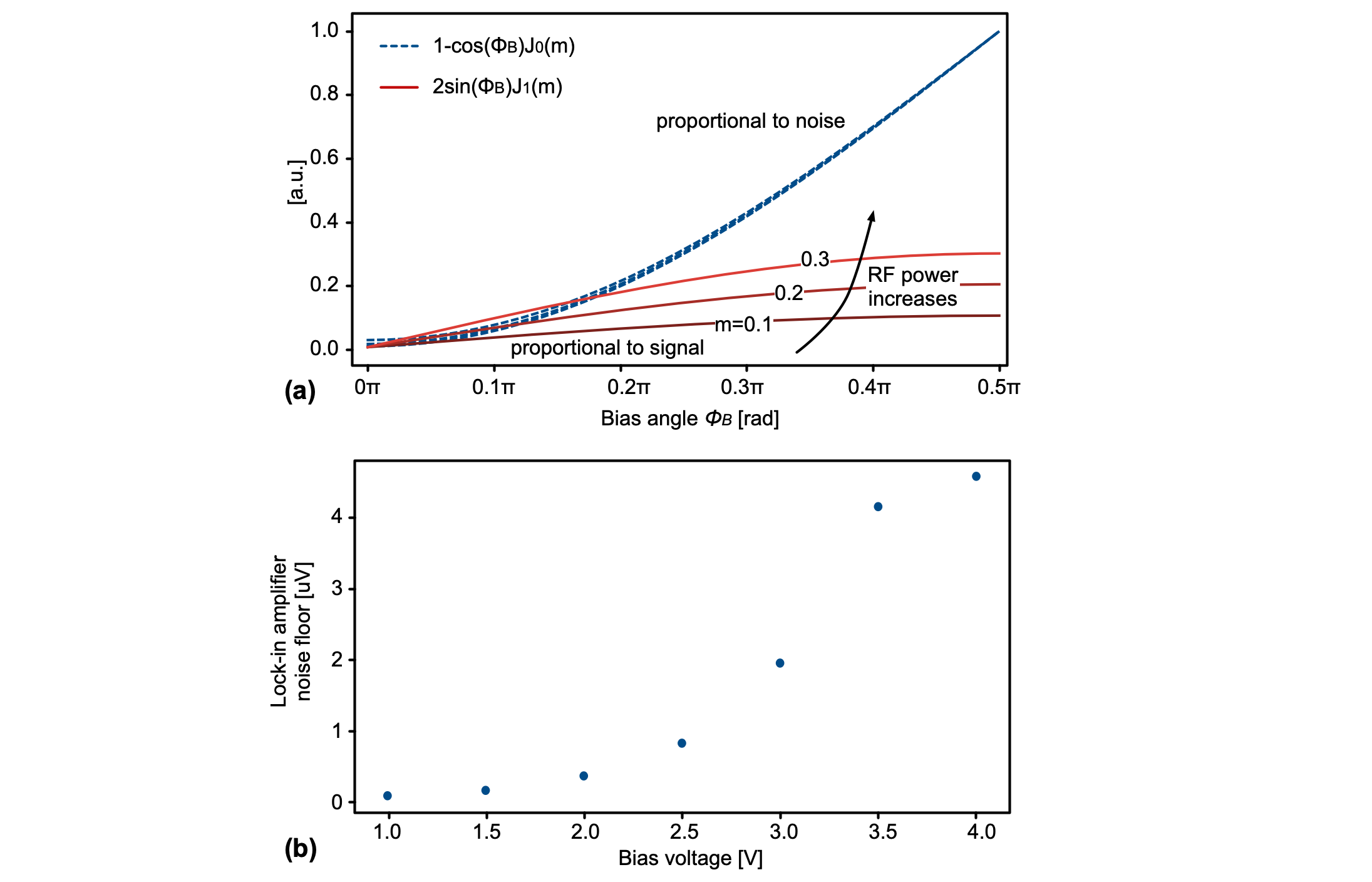}
\caption{(a) The DC part and the modulation part of the optical power at different DC bias points, which contribute to the noise floor and the signal in the lock-in amplifier measurement, respectively; (b) Measured noise floor at the lock-in amplifier at various bias voltages.}
\label{fig:modulation}
\end{figure}

The influence of the DC bias voltages and modulation depths can be investigated with \eqref{eq:power} and \eqref{eq:trans-probe}. As shown in \eqref{eq:trans-probe}, only the beating between the AC terms in \eqref{eq:power} will be detected by the lock-in amplifier. The DC terms in \eqref{eq:power}, which represents the average optical power, primarily contributes to the noise received at the photodiode \cite{Marpaung2009}. Fig.~\ref{fig:modulation} (a) depicts the AC and DC terms in \eqref{eq:power} at various bias angles. With a constant bias, increasing the modulation depth, i.e., increasing the RF power, would increase the signal strength. Conversely, tuning the bias angle affects both the signal strength and the average optical power. At the quadrature bias angle ($\Phi_B = 0.5\pi$), the AC term is the highest, however, it also leads to a high DC term. Shifting the DC bias towards the null point ($\Phi_B = 0$) would reduce both the signal and the noise strength. Nevertheless, the DC term drops quicker than the AC term, which potentially leads to a higher SNR. 

The optimized configuration of the bias voltage applied to the intensity modulators is a trade-off between the signal strength and the noise floor level. To keep the noise floor low while maintaining the signal strength, the DC bias point in the experiments should be kept between 0 and $0.5\pi$, which is similar to the low-biasing technique widely used in the microwave photonics field \cite{Daulay2022}. Fig.~\ref{fig:modulation}~(b) shows the measured noise floor at the lock-in amplifier at various bias voltages. In the experiments, we put the bias voltage between 2.0 and 2.5~V for the best SNR. However, notice that the drifting of the modulator might result in different optimized bias voltage. 

\subsection{Influences of modulation frequencies and detected frequency}
\begin{figure}[t!]
\centering
\includegraphics[width=\linewidth]{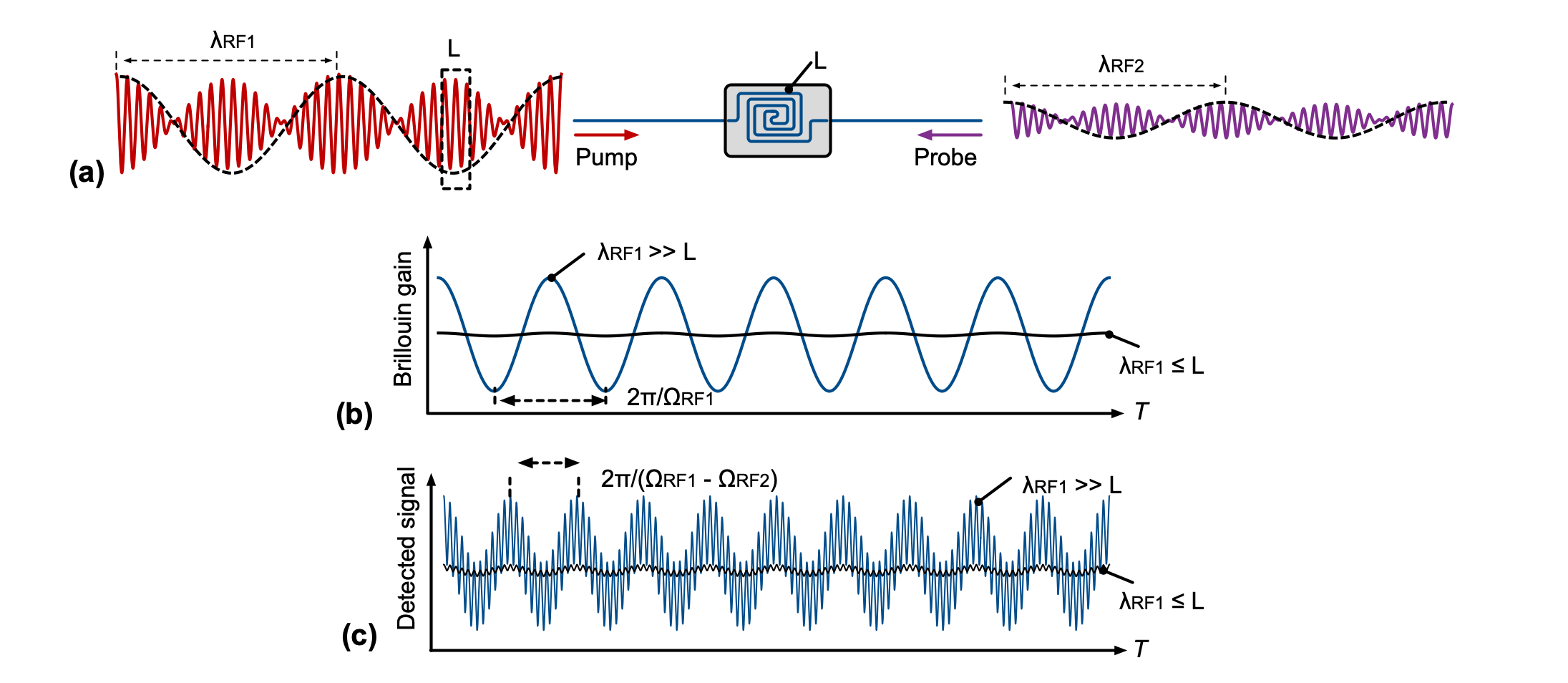}
\caption{(a) Interaction between the intensity-modulated pump and probe when the waveguide length $L$ is much shorter than the modulation envelop $\lambda_{\rm RF1}$. (b) Brillouin gain from the waveguide when $\lambda_{\rm RF1}$ is much larger than the waveguide length $L$  or  $\lambda_{\rm RF1}$ is smaller or comparable to $L$. (c) Detected signal strength at the lock-in amplifier in these two situation.}
\label{fig:mod-frequency}
\end{figure}
The modulation frequencies of the pump and probe would influence the strength of the detected signal at the lock-in amplifier. Fig.~\ref{fig:mod-frequency}(a) illustrates the interaction between the intensity-modulated pump and probe inside the waveguide. The optical wavelength (in the range of micrometers) is much shorter than the waveguide (in the range of centimeters). However, the envelope of the pump $\lambda_{\rm RF1}$, which is determined by the modulation frequency, might have a wavelength that is comparable to the waveguide length. As a result, the modulation frequency of the pump might influence the total SBS gain.

Two extreme scenarios are considered in Fig.~\ref{fig:mod-frequency} (b). When $\lambda_{\rm RF1}$ is much longer than the length of the waveguide $L$, i.e., when the modulation frequency is low enough, the pump strength inside the waveguide would be different at any given time, depending on the phase of the pump envelopes. In other words, the SBS gain from the waveguide will oscillate at the modulation frequency. Conversely, when $\lambda_{\rm RF1}$ is comparable to or even shorter than the spiral length, the pump power inside the waveguide would be almost constant, leading to a diminished amplitude in the SBS gain oscillation. The influence of the modulation frequency on the SBS gain is also reflected in the integration term in \eqref{eq:trans-probe}.

Since the lock-in amplifier detects the signal at the beating frequency of the pump and probe, lower SBS gain oscillation amplitude would lead to a lower amplitude in the detected signal, as shown in Fig.~\ref{fig:mod-frequency} (c). Consequently, from the signal strength point of view, a relatively lower modulation frequency is preferred in the experiment.

Furthermore, an inappropriate modulation frequency may lead to a potentially inaccurate evaluation of the Brillouin gain coefficient. Typically, the Brillouin gain coefficient of the waveguide is determined by comparing the SBS signal from the fiber and waveguide peak. Taking into account the influence of the modulation frequency, the Brillouin gain coefficient can be calculated as:

\begin{equation}
g_{\rm{B}, \rm{wg}} = \frac{V_{\rm{wg}}}{V_{\rm{fiber}}}\frac{g_{\rm{B}, \rm{fiber}}P_{\rm{pump,fiber}}\int_0^{L_{\rm fiber}}{\rm cos}\left(\frac{2\pi}{\lambda_{\rm RF1}}z\right)dz}{P_{\rm{pump,wg}}\int_0^{L_{\rm wg}}{\rm cos}\left(\frac{2\pi}{\lambda_{\rm RF1}}z \right) e^{-\alpha z}dz}
\label{eq:comp2}
\end{equation}

When the modulation frequency $\Omega_{\rm RF1}$ is too high (in the order of tens of GHz), i.e., the wavelength of the envelope $\lambda_{\rm RF1}$ is comparable to the waveguide length, the detected SBS signals from both the fiber and the waveguide would be small, and the SNR of the measured signal would be poor. When $\lambda_{\rm RF1}$ is larger than the length of the waveguide but not the fiber, only the fiber peak would be diminished, and the calculation based on \eqref{eq:comp} would overestimate the Brillouin gain coefficient of the waveguide. When $\lambda_{\rm RF1}$ is substantially larger than the length of both the waveguide and the fiber, the expression in \eqref{eq:comp2} converges to \eqref{eq:comp}, and the detected signals will also be the strongest. 

However, applying an excessively low modulation frequency is also discouraged. Given that EDFA is applied to amplify the intensity-modulated pump and probe, the amplified spontaneous emission (ASE) from the EDFA will significantly increase when the modulation frequency is too low, leading to a rise in the noise floor. In our experiment, we set the the modulation frequencies at 10~MHz and 10.075~MHz for the pump and probe respectively, corresponding to $\lambda_{\rm RF1}$ of approximately 30~m. This value is much longer than both the waveguide (3~cm) and the fiber (5~m) lengths while the ASE from the EDFA remains negligible. 

\begin{figure*}[t!]
\centering
\includegraphics[width=\linewidth]{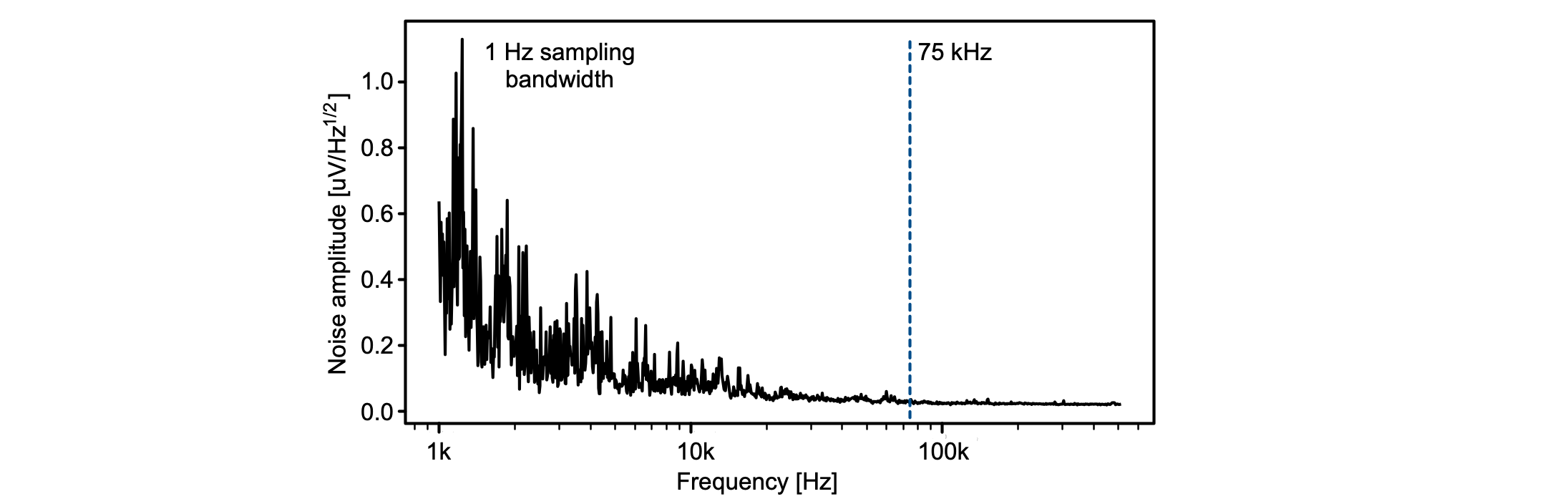}
\caption{Measured input-referred noise spectrum of the lock-in amplifier setup.}
\label{fig:noise}
\end{figure*}

The selection of the detected signal frequency is mainly related to the technical noise in the setup. Fig.~\ref{fig:noise} shows the measured input-referred noise spectrum of the lock-in amplifier setup. While the noise at low frequencies is relatively high, the measured noise is close to the white noise floor from 75~kHz. By modulating the pump and the probe signal at slightly different frequencies (10~MHz and 10~MHz + 75 kHz in our experiment), this configuration shifts the detected frequency from DC to 75~kHz, significantly reducing the influences of technical noise at low frequencies.

\subsection{Influences of the facet reflection and the Kerr effect}
For samples with high reflection at the facet, strong oscillation would appear in the noise floor in the dual-intensity-modulated pump-probe lock-in amplifier setup. This oscillation is caused by the mixing between the reflected pump, which is comparable or even higher than the probe, and the transmitted probe. As shown in Fig.~\ref{fig:double}(c), the oscillation frequency in the noise floor is close to the  free spectral range (FSR) of the Fabry-Perot (FP) cavity formed by the two facets of the sample, indicating that the transmitted spectra of FP cavity is imprinted into the noise floor. Consequently, to reduce the influence of the oscillation in the noise floor, samples with anti-reflection coating at the facets are preferred. Moreover, high-rejection optical bandpass filter can also be introduced to filter out the reflected pump.

\begin{figure}[t!]
\centering
\includegraphics[width=\linewidth]{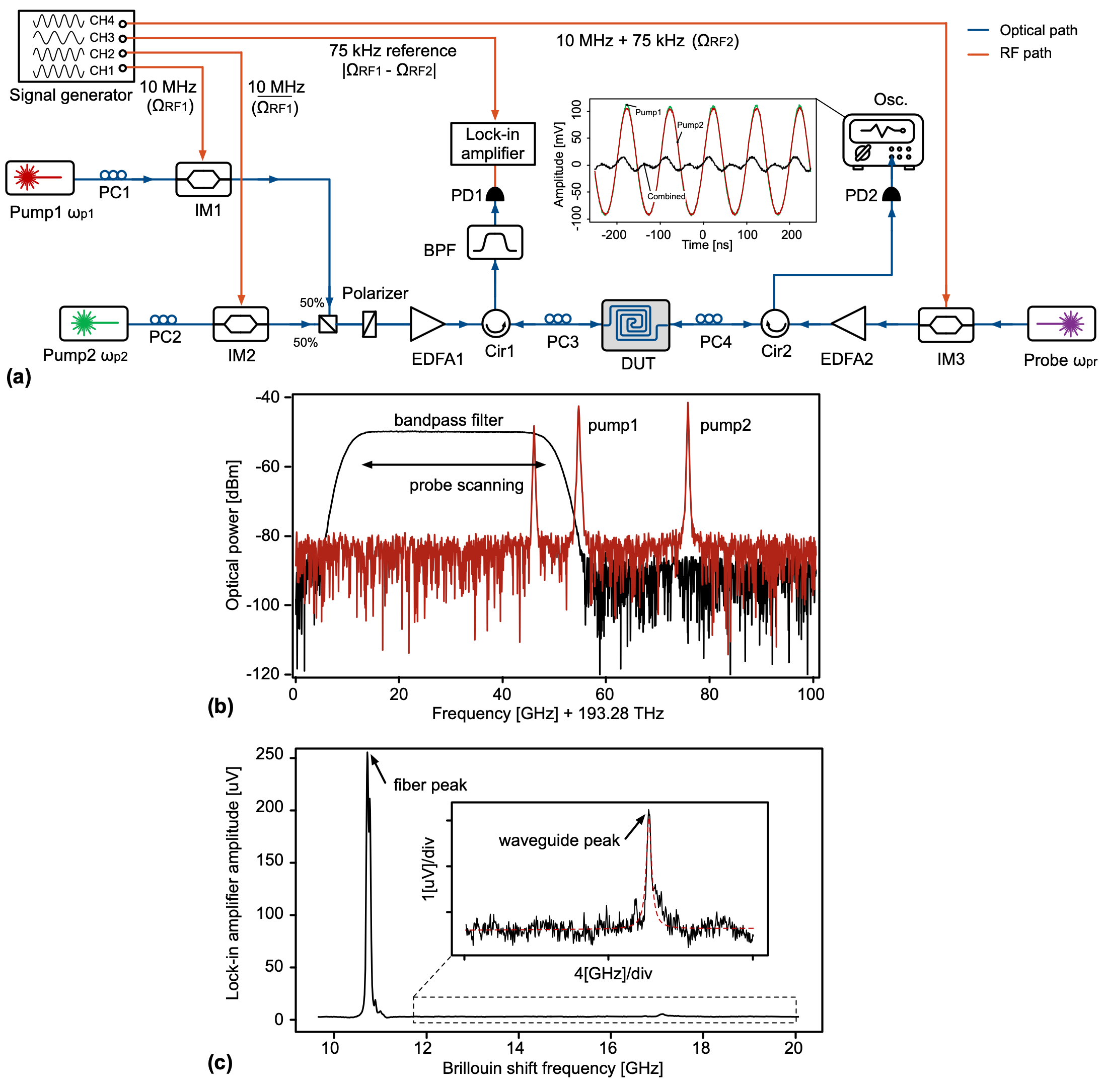}
\caption{(a) Schematic of the triple-intensity-modulated lock-in amplifier setup; (b) The optical spectra of the two pumps, the probe, and the optical bandpass filter; (c) The measured SBS signals from the 3~cm long SiON waveguide and the 5~m long optical fiber in the setup. The noise floor is lower compared to the double-intensity modulation setup and the waveguide peak appears at 17.15 GHz.}
\label{fig:triple}
\end{figure}

The noise floor oscillation is also caused by the Kerr effect in the waveguide. Due to the Kerr effect, the FSR of the cavity would change as a function of time:

\begin{equation}
    \frac{d FSR}{d t} = \frac{d FSR}{d P_{p}} \cdot \frac{d P_{p}}{d t} 
    \approx \frac{2\pi n_2 c}{n_g^2LA_{\rm eff}} \cdot \left[P_{p0}\Omega_{RF1}{\rm sin}\phi_{B1}J_1{\rm sin}(\Omega_{RF1}t)\right]
    \label{eq:fsr}
\end{equation}
in which $n_2$ is nonlinear index,  $c$ is the speed of light in vacuum, $n_g$ is group index, $L$ is the waveguide length, and $A_{\rm eff}$ is the effective optical mode area.

As indicated by \eqref{eq:fsr}, the Kerr effect imparts the pump oscillation onto the FSR of the cavity. This FSR oscillation affects the transmitted probe, giving rise to spurious signals at the beating frequency $|\Omega_{RF1} - \Omega_{RF2}|$. In other words, the Kerr effect also contributes to the noise floor oscillation in the intensity-modulated pump-probe lock-in amplifier setup. To alleviate the influence of the Kerr effect, the pump variation needs to be reduced, which can be realized with the triple-intensity-modulated pump-probe technique.

\subsection{Triple-intensity modulation pump-probe lock-in amplifier setup}

Fig.~\ref{fig:triple}~(a) shows the schematic of the triple-intensity-modulated pump-probe lock-in amplifier setup. The core of this setup is the same as the double-intensity modulation setup, with the addition of an auxiliary Pump~2 (1561.154~nm). Pump~2 is modulated by an RF signal ($\overline{\Omega_{\rm RF1}}$) at the same frequency but with an inverse phase compared to the RF signal applied to Pump~1. Pump~1 and Pump~2 are combined with a 50:50 splitter and are sent through a polarizer before being amplified by EDFA1. This dual-pump configuration cancels out the pump power fluctuation inside the waveguide, as shown in the inset of Fig.~\ref{fig:triple}~(a). As indicated by \eqref{eq:fsr}, the influence of the Kerr effect can be minimized consequently. The spectrum configurations of the two pumps, the probe, and the optical bandpass filter are shown in Fig.~\ref{fig:triple}~(b). Pump~2 operates at a considerably higher frequency than Pump~1, ensuring the generated SBS signal from Pump~2 is at the stopband of the filter and would not influence the determination of the Brillouin frequency shift. 

The measurement result with the triple-intensity-modulated pump-probe lock-in amplifier setup is presented in Fig.~\ref{fig:triple}~(c). The noise floor is much lower compared to the double-intensity modulation. Moreover, the oscillation in the noise floor is negligible. The SBS response from the SiON waveguide is clearly visible at a Brillouin frequency shift of 17.15~GHz with a Brillouin gain coefficient of 0.75~m$^{-1}$W$^{-1}$.

\section{Discussions}
\begin{table}[t!]
\centering
\caption{Comparison between different SiON platforms}
\resizebox{\textwidth}{!}{\begin{tabular}{cccccc}
\hline
\multirow{2}{*}{} &
  \multirow{2}{*}{\begin{tabular}[c]{@{}c@{}}Refractive \\ index\end{tabular}} &
  \multirow{2}{*}{\begin{tabular}[c]{@{}c@{}}Propagation\\ loss {[}dB/cm{]}\end{tabular}} &
  \multirow{2}{*}{\begin{tabular}[c]{@{}c@{}}Brillouin gain\\ coefficient {[}m$^{-1}$W$^{-1}${]}\end{tabular}} &
  \multirow{2}{*}{\begin{tabular}[c]{@{}c@{}}Brillouin frequency \\ shift {[}GHz{]}\end{tabular}} & \multirow{2}{*}{\begin{tabular}[c]{@{}c@{}}Linewidth\\ {[}MHz{]}\end{tabular}} \\
          &  &  &  & & \\ \hline
\cite{Ye2023}  & 1.51 & 0.25 & 0.32 & 14.48 & 105 \\ \hline
\cite{Zerbib2023}  & 1.70 & 0.1 & 0.09 & 16 & 350 \\ \hline
This work & 1.65 & 0.8 & 0.75 & 17.15 & 177\\ \hline
\end{tabular}}
\label{tab:comp}
\end{table}


Table~\ref{tab:comp} summarizes the Brillouin nonlinearities of the SiON platform in this work, alongside two additional SiON platforms in the literature \cite{Ye2023,Zerbib2023}. Varied refractive indexes, in other words, different O/N ratios, result in distinct Brillouin nonlinearities, confirming the tunability of the SBS responses in SiON.  Despite the refractive index of the SiON in this study falling between those reported in \cite{Ye2023} and \cite{Zerbib2023}, the Brillouin nonlinearity of this platform, particularly the Brillouin gain coefficient and Brillouin frequency shift, deviates from the values presented in \cite{Ye2023} and \cite{Zerbib2023}. This deviation can be attributed to the diverse mechanical properties of SiON. Considering that the Brillouin nonlinearity is intricately linked to both the optical and mechanical properties of the material. While optical properties are primarily dictated by stoichiometry, mechanical properties significantly depend on the fabrication process \cite{kalb2022}. Variations in mechanical properties, i.e., Young's modulus, density, and Poisson ratio, can lead to substantial disparities in Brillouin nonlinearities. Consequently, a comprehensive exploration of the mechanical properties of SiON is imperative for further Brillouin nonlinearity characterizations in SiON platforms.

From the characterization perspective, this work discusses various on-chip Brillouin nonlinearity characterization techniques, with a particular emphasis on a detailed analysis of the intensity-modulated pump-probe lock-in amplifier technique. Employing the triple-intensity-modulated pump-probe setup with the optimal configurations, we successfully detected the SBS responses from SiON waveguides with a gain below 0.1~dB. This intensity-modulated pump-probe lock-in amplifier setup proves to be well-suited for Brillouin nonlinearity characterizations of platforms with comparable SBS gain, such as silicon nitride, or aluminum oxide. However, for platforms exhibiting substantially stronger Brillouin nonlinearity, as observed in tellurite-covered silicon nitride \cite{roel2023tellurite}, chalcogenide \cite{neijts2023onchip}, or thin-film lithium niobate \cite{kaixuan2023tfln}, alternative techniques can also be considered.

\begin{backmatter}
\bmsection{Funding}
The authors acknowledge funding from the European Research Council Consolidator Grant (101043229 TRIFFIC) and Nederlandse Organisatie voor Wetenschappelijk Onderzoek (NWO) Start Up (740.018.021) and EC within the Horizon 2020 project EPIQUS (grant 899368).

\bmsection{Disclosures}
The authors declare no conflicts of interest.

\bmsection{Data Availability Statement}
The data that support the findings of this study are available from the corresponding authors upon reasonable request.
\end{backmatter}


\bibliography{library}

\end{document}